%% file: main.tex
\documentclass[a4paper,11pt]{article}

\usepackage{jcappub} 

\usepackage{graphicx}
\usepackage{amsmath,amsfonts,amssymb}
\usepackage{txfonts}
\usepackage{color}
\usepackage{float}
\usepackage{dblfloatfix}
\usepackage{afterpage}
\usepackage{ifthen}
\usepackage[morefloats=12]{morefloats}
\usepackage{placeins}
\usepackage{multicol}
\usepackage[switch]{lineno}
\definecolor{linkcolor}{rgb}{0.6,0,0}
\definecolor{citecolor}{rgb}{0,0,0.75}
\definecolor{urlcolor}{rgb}{0.12,0.46,0.7}
\hypersetup{linktocpage}
\usepackage{bold-extra}
\usepackage{xcolor}
\usepackage{tabularx}
\usepackage{csquotes}
\usepackage{breakurl}
\usepackage{lipsum}
\usepackage{mathtools}
\usepackage{cuted}
\usepackage{upgreek}
\usepackage{lineno}

\usepackage[nameinlink,noabbrev]{cleveref}

\input{Planck}

\def\WMAP{{\it WMAP}}
\def\COBE{{\it COBE}}
\def\DIRBE{{DIRBE}}

\def\commanderthree{\texttt{Commander3}}

\let\vec\mathbf
\renewcommand{\d}[0]{\vec{d}}

\newcommand{\n}[0]{\vec{n}}

\newcommand{\s}[0]{\vec{s}}
\renewcommand{\a}[0]{\vec{a}}

\renewcommand{\L}[0]{\tens{L}}

\newcommand{\BP}{\textsc{BeyondPlanck}}

\newcommand{\lbs}[0]{LBS}
\def\LB{\textit{LiteBIRD}}

\def\inv{^{-1}}

\title{\bfseries On the computational feasibility of Bayesian end-to-end analysis of \LB\ simulations within Cosmoglobe} 
\makeatletter
\def\@xfootnote[#1]{%
  \protected@xdef\@thefnmark{#1}%
  \@footnotemark\@footnotetext}
\makeatother
\input{common/authors_affiliations_LB_2022.01_general_jcap.tex}
\emailAdd{unnif@astro.uio.no}

\abstract{We assess the computational feasibility of end-to-end Bayesian analysis of the JAXA-led \LB\ experiment by analysing simulated time ordered data (TOD) for a subset of detectors through the Cosmoglobe and  \commanderthree\  framework. The data volume for the simulated TOD is 1.55\,TB, or 470\,GB after Huffman compression. From this we estimate a total data volume of 238\,TB for the full three year mission, or 70\,TB after Huffman compression. We further estimate the running time for one Gibbs sample, from TOD to cosmological parameters, to be approximately 3000\,CPU\,hours. The current simulations are based on an ideal instrument model, only including correlated $1/f$ noise. Future work will consider realistic systematics with full end-to-end error propagation. We conclude that these requirements are well within capabilities of future high-performance computing systems.}


\begin{document}

\maketitle
\flushbottom

\input{Intro}
\input{LiteBIRD}

\input{Algorithm}

\input{Simulations}

\input{Results}

\input{Conlusions}

\input{common/Acknowledgments}  

\bibliographystyle{JHEP}

\bibliography{bibliography}

\end{document}

%% file: Planck.tex
\def\setsymbol#1#2{\expandafter\def\csname #1\endcsname{#2}}
\def\getsymbol#1{\csname #1\endcsname}

\def\Planck{\textit{Planck}}

\newbox\tablebox    \newdimen\tablewidth
\def\leaderfil{\leaders\hbox to 5pt{\hss.\hss}\hfil}
\def\endPlancktablewide{\tablewidth=\textwidth 
    $$\hss\copy\tablebox\hss$$
    \vskip-\lastskip\vskip -2pt}
\def\tablenote#1 #2\par{\begingroup \parindent=0.8em
    \abovedisplayshortskip=0pt\belowdisplayshortskip=0pt
    \noindent
    $$\hss\vbox{\hsize\tablewidth \hangindent=\parindent \hangafter=1 \noindent
    \hbox to \parindent{$^#1$\hss}\strut#2\strut\par}\hss$$
    \endgroup}
\def\doubleline{\vskip 3pt\hrule \vskip 1.5pt \hrule \vskip 5pt}

\def\L2{\ifmmode L_2\else $L_2$\fi}

\def\DeltaT{\ifmmode \Delta T\else $\Delta T$\fi}
\def\deltat{\ifmmode \Delta t\else $\Delta t$\fi}
\def\fknee{\ifmmode f_{\rm knee}\else $f_{\rm knee}$\fi}
\def\Fmax{\ifmmode F_{\rm max}\else $F_{\rm max}$\fi}
\def\solar{\ifmmode{\rm M}_{\mathord\odot}\else${\rm M}_{\mathord\odot}$\fi}
\def\Msolar{\ifmmode{\rm M}_{\mathord\odot}\else${\rm M}_{\mathord\odot}$\fi}
\def\Lsolar{\ifmmode{\rm L}_{\mathord\odot}\else${\rm L}_{\mathord\odot}$\fi}
\def\inv{\ifmmode^{-1}\else$^{-1}$\fi}
\def\mo{\ifmmode^{-1}\else$^{-1}$\fi}
\def\sup#1{\ifmmode ^{\rm #1}\else $^{\rm #1}$\fi}
\def\expo#1{\ifmmode \times 10^{#1}\else $\times 10^{#1}$\fi}
\def\,{\thinspace}
\def\lsim{\mathrel{\raise .4ex\hbox{\rlap{$<$}\lower 1.2ex\hbox{$\sim$}}}}
\def\gsim{\mathrel{\raise .4ex\hbox{\rlap{$>$}\lower 1.2ex\hbox{$\sim$}}}}

\def\simprop{\mathrel{\raise .4ex\hbox{\rlap{$\propto$}\lower 1.2ex\hbox{$\sim$}}}}
\def\deg{\ifmmode^\circ\else$^\circ$\fi}
\def\pdeg{\ifmmode $\setbox0=\hbox{$^{\circ}$}\rlap{\hskip.11\wd0 .}$^{\circ}
          \else \setbox0=\hbox{$^{\circ}$}\rlap{\hskip.11\wd0 .}$^{\circ}$\fi}
\def\arcs{\ifmmode {^{\scriptstyle\prime\prime}}
          \else $^{\scriptstyle\prime\prime}$\fi}
\def\arcm{\ifmmode {^{\scriptstyle\prime}}
          \else $^{\scriptstyle\prime}$\fi}
\newdimen\sa  \newdimen\sb
\def\parcs{\sa=.07em \sb=.03em
     \ifmmode \hbox{\rlap{.}}^{\scriptstyle\prime\kern -\sb\prime}\hbox{\kern -\sa}
     \else \rlap{.}$^{\scriptstyle\prime\kern -\sb\prime}$\kern -\sa\fi}
\def\parcm{\sa=.08em \sb=.03em
     \ifmmode \hbox{\rlap{.}\kern\sa}^{\scriptstyle\prime}\hbox{\kern-\sb}
     \else \rlap{.}\kern\sa$^{\scriptstyle\prime}$\kern-\sb\fi}
\def\ra[#1 #2 #3.#4]{#1\sup{h}#2\sup{m}#3\sup{s}\llap.#4}
\def\dec[#1 #2 #3.#4]{#1\deg#2\arcm#3\arcs\llap.#4}
\def\deco[#1 #2 #3]{#1\deg#2\arcm#3\arcs}
\def\rra[#1 #2]{#1\sup{h}#2\sup{m}}

\def\dots{\relax\ifmmode \ldots\else $\ldots$\fi}
\def\WHzsr{\ifmmode $W\,Hz\mo\,sr\mo$\else W\,Hz\mo\,sr\mo\fi}
\def\mHz{\ifmmode $\,mHz$\else \,mHz\fi}
\def\GHz{\ifmmode $\,GHz$\else \,GHz\fi}
\def\mKs{\ifmmode $\,mK\,s$^{1/2}\else \,mK\,s$^{1/2}$\fi}
\def\muKs{\ifmmode \,\mu$K\,s$^{1/2}\else \,$\mu$K\,s$^{1/2}$\fi}
\def\muKRJs{\ifmmode \,\mu$K$_{\rm RJ}$\,s$^{1/2}\else \,$\mu$K$_{\rm RJ}$\,s$^{1/2}$\fi}
\def\muKHz{\ifmmode \,\mu$K\,Hz$^{-1/2}\else \,$\mu$K\,Hz$^{-1/2}$\fi}
\def\MJysr{\ifmmode \,$MJy\,sr\mo$\else \,MJy\,sr\mo\fi}
\def\MJysrmK{\ifmmode \,$MJy\,sr\mo$\,mK$_{\rm CMB}\mo\else \,MJy\,sr\mo\,mK$_{\rm CMB}\mo$\fi}
\def\microns{\ifmmode \,\mu$m$\else \,$\mu$m\fi}

\def\muK{\ifmmode \,\mu$K$\else \,$\mu$\hbox{K}\fi}
\def\microK{\ifmmode \,\mu$K$\else \,$\mu$\hbox{K}\fi}
\def\muW{\ifmmode \,\mu$W$\else \,$\mu$\hbox{W}\fi}
\def\kms{\ifmmode $\,km\,s$^{-1}\else \,km\,s$^{-1}$\fi}
\def\kmsMpc{\ifmmode $\,\kms\,Mpc\mo$\else \,\kms\,Mpc\mo\fi}
\providecommand{\sorthelp}[1]{}

%% file: common/authors_affiliations_LB_2022.01_general_jcap.tex
\author[1]{R.\,Aurvik,}
\author[1]{M.\,Galloway,}
\author[1]{E.\,Gjerløw,}
\author[1]{U.\,Fuskeland,\footnote[*]{Corresponding Author.}}
\author[1]{A.\,Basyrov,}
\author[2,3]{M.\,Bortolami,}
\author[1]{M.\,Brilenkov,}
\author[3,4,5]{P.\,Campeti,}
\author[1]{H.\,K.\,Eriksen,}
\author[6,7]{L.\,T.\,Hergt,}
% ...
\affiliation[1]{Institute of Theoretical Astrophysics, University of Oslo, Blindern, Oslo, Norway}
\affiliation[2]{Dipartimento di Fisica e Scienze della Terra, Università di Ferrara, Via Saragat 1, 44122 Ferrara, Italy}
\affiliation[3]{INFN Sezione di Ferrara, Via Saragat 1, 44122 Ferrara, Italy}
\affiliation[4]{Max Planck Institute for Astrophysics, Karl-Schwarzschild-Str. 1, D-85748 Garching, Germany}
\affiliation[5]{Excellence Cluster ORIGINS, Boltzmannstr. 2, 85748 Garching, Germany}
\affiliation[6]{Department of Physics and Astronomy, University of British Columbia, 6224 Agricultural Road, Vancouver, BC V6T1Z1, Canada}
\affiliation[7]{Université Paris-Saclay, CNRS/IN2P3, IJCLab, 91405 Orsay, France}
% ...

%% file: Intro.tex
\section{Introduction}
\label{sec:introduction}

The $\Lambda$CDM model is the most successful model describing the Universe as we know it, with massive observational evidence \citep{Komatsu:2014ioa, planck2016-l09, planck2016-l06, planck2016-l10} supporting its predictions. One feature of this theory is an early period of cosmic inflation during which the Universe expanded exponentially \citep{Guth_1981, Sato:1980yn, Albrecht:1982wi, Linde:1981mu, Grishchuk:1974ny, Starobinski_1979}. Fluctuations in the inflaton field would generate density perturbations, becoming the structures we see in the Universe today. An additional prediction is that tensor perturbations, or gravitational waves, would be generated during this early inflationary period. Such tensor perturbations would produce a curl pattern in the Cosmic Microwave Background (CMB), known as $B$ modes \citep{Kamionkowski:1996zd, Seljak:1996gy}. The detection of these $B$ modes would be additional and definitive evidence for inflation, but due to their low signal strength pushes the requirements of the instruments, sensitivities, and data processing to extremely high levels.

The CMB radiation was released during the period of recombination, 380\,000 years after the beginning of our Universe. Today, after the Universe has expanded for 13.8\,billion years, this radiation has cooled to 2.7255\,K~\citep{fixsen2009}. After the first detection in the microwave regime in 1965 by ref.~\cite{penzias:1965}, there have been several CMB experiments, three of which were full-sky space missions. The latest state-of-the-art mission was the \Planck\ space mission launched in 2009 \citep{planck2016-l01}, following up on the Cosmic Background Explorer (\COBE) launched in 1989 \citep{mather:1994,smoot:1992} and the Wilkinson Microwave Anisotropy Probe (\WMAP) launched in 2001 \citep{bennett2012}. \COBE, \WMAP, and \Planck\ have all been immensely successful; \COBE\ measuring the blackbody spectrum of the CMB monopole \citep{mather:1994} and the temperature fluctuations of the CMB across the sky \citep{smoot:1992}, \WMAP\ giving us the age of the Universe and reducing the allowed $\Lambda$CDM parameter space by a factor of 68\,000, and \Planck\ giving us cosmic variance limited observations in intensity and a sub-percent precision on all $\Lambda$CDM parameters except for the optical depth due to reionisation, $\tau$. 

However, the \WMAP\ and \Planck\ polarisation observations did not result in a detection of the predicted $B$-mode signal. The current upper limit of the tensor-to-scalar ratio $r$ used to quantify the strength of the $B$-mode signal is constrained by the BICEP/Keck 2018 data at $r<0.036$ \citep{bicep2018}, decreasing to $0.032$ when combined with the latest \Planck\ PR4 data \citep{Tristram_2022}. For reference, some single-field slow-roll inflationary theories predict $r\simeq0.003$ \citep{Kamionkowski:2015yta}. In order to set tighter constraints on $r$, the sensitivity of future experiments needs to improve by an order of magnitude, which sets far more stringent requirements on the sensitivity and precision of future instruments and data analysis pipelines. 

One of the next-generation CMB experiments is the Japanese led Lite (Light) satellite for the studies of $B$-mode polarisation and Inflation from cosmic background Radiation Detection (\LB), due to be launched in the early 2030's \citep{LB_PTEP}. \LB's main goal is the detection of the inflationary $B$-mode signal with uncertainty of $\delta r \lessapprox 0.001$ \citep{Suzuki_2018}. Its effective sensitivity will be $2\,\mathrm{\mu K\,arcmin}$, corresponding to the combined integration time of nearly 1000 \Planck\ missions \citep{LB_PTEP}. 

To process these data robustly and efficiently, the analysis pipelines required for the \LB\ data will need to be significantly improved compared to the \Planck\ era. A major lesson learned from \Planck\ was that the ultimate limiting factor in terms of overall sensitivity is the interplay between systematic uncertainties in the instrument and astrophysical sky models \citep{planck2016-l02}. \LB's much lower noise level will make it even more sensitive to systematic effects from the instruments than \Planck\ was, and this underscores the need for robust data analysis pipelines that simultaneously account for all relevant effects to ensure a clean detection of the $B$-mode signal.

The \BP\ collaboration developed a tool for global integrated end-to-end Bayesian data analysis, called \commanderthree\ \citep{bp03}. This framework implements a comprehensive Gibbs sampler starting from time ordered data (TOD), and outputting high-level astrophysical and cosmological products, including cosmological parameters. This framework is sufficiently efficient to allow thousands of iterations through TOD calibrations, map-making, and component separation, and the resulting Monte Carlo samples map out a full joint posterior distribution. These samples can thereby replace traditional frequentist simulations for error propagation \citep{bp04}. This framework was demonstrated with the \Planck\ Low Frequency Instrument (LFI) data in ref.~\citep{bp01} and companion papers, resulting in new state-of-the-art LFI frequency maps between 30 and 70\,GHz.

The Cosmoglobe\footnote{https://cosmoglobe.uio.no} project aims to apply this global analysis process to as many radio, microwave, and sub-millimeter data sets as possible within an Open Source and supportive community. The ultimate goal of Cosmoglobe is thus to build a global joint model of the microwave sky, while properly accounting for as many instrument-specific systematic effects as possible. As a first step, ref.~\citep{watts2023cosmoglobe} jointly analysed \WMAP\ and \Planck\ LFI TOD in this framework. By using information from one data set to break degeneracies in the other, the frequency maps of both data sets were significantly improved. Particularly noteworthy was that the \WMAP\ $W$-band polarisation sky map was excluded in the official \WMAP\ cosmological analysis due to low $\ell$ systematics \citep{jarosik2010}, but after the analysis done by ref.~\citep{watts2023cosmoglobe} the improved map was, for the first time, consistent with all other comparable maps. Recently, the same framework has also been used to re-process the \COBE/\DIRBE\ data, resulting in improved frequency maps \citep{CG02_01}, stronger CIB monopole constraints \citep{CG02_03}, a better zodiacal light model \citep{CG02_02}, a deeper starlight model \citep{CG02_04}, and a novel three-component thermal dust model \citep{CG02_05}. 

While these analyses have proved that the end-to-end joint analysis approach is indeed feasible for currently available data sets, it is not obvious whether the same approach can be scaled to the much higher computational demands of next-generation experiments. In this work, we therefore extend the Cosmoglobe \commanderthree\ code to support future \LB\ measurements, and apply this to the first generation of simulated \LB\ TOD generated by the \LB\ collaboration's newly developed simulation pipeline. The main goal of this analysis is to assess the computational feasibility of global Bayesian analysis for \LB\ data. Specifically, we estimate the total computational cost of analysing the full 3-year mission with the full focal plane by extrapolating actual run times from a reduced data volume spanning one year of data and a subset of detectors.

For this analysis, we used the TOD produced by the end-to-end pipeline described in ref.~\citep{bortolami:2025}. This pipeline simulates the operations of the three instruments onboard the \LB\ spacecraft and saves the TOD in HDF5 files, which we then convert to be compatible with the established \commanderthree\ data format.

The rest of the paper is organised as follows. First, we give a brief introduction to the \LB\ experiment in \cref{sec:LB} before describing the algorithm in \cref{sec:Algorithm}. Next, the production of the time-ordered \LB\ simulations is explained in \cref{sec:sims}. Finally, our results are presented in \cref{sec:Results}, and conclusions are drawn in \cref{sec:Conclusions}.

%% file: LiteBIRD.tex
\section{The \LB\ experiment}
\label{sec:LB}
\LB\ is an experiment led by the Japan Aerospace Exploration Agency (JAXA), with contributors in both Europe and North America. \LB\ is to be launched to the second Lagrange point of the Sun-Earth system in the early 2030's. There it will map the microwave sky for three years to produce the cleanest full-sky map of the polarised sky available. For a full overview of the \LB\ experiment we refer the interested reader to ref.~\citep{LB_PTEP}. 

\LB's main scientific goal is to measure the CMB polarisation to probe the inflationary $B$-mode signal. It aims at either making a discovery or ruling out large classes of currently viable inflationary models \citep[e.g. ref.][]{Linde_2017}. \LB\ also has several secondary scientific goals, including a greater understanding of the reionisation of the Universe, detection of cosmic birefringence, and detecting primordial magnetic fields. 

The current \LB\ baseline design implements a total of 4508 bolometers distributed over 15 frequency bands between 34 and 448\,GHz, with focal planes cooled to 0.1\,K to have low thermal disturbances \citep{LB_PTEP}. It consists of three different telescopes: the Low-Frequency Telescope (LFT; 34--161\,GHz; \citealp{Sekimoto_2020}), the Mid-Frequency Telescope (MFT; 89--228\,GHz) and the High-Frequency Telescope (HFT; 166--448\,GHz; \citealp{Montier_2020}). Some frequency bands are divided across two different telescopes to have a partial overlap between telescopes, and thus increase the ability to monitor systematic effects within a given frequency. The telescopes also includes a levitating continuously spinning half-wave plates (HWP) to improve polarisation angle coverage and decrease the instrumental $1/f$ noise \citep{Montier_2020}. 

The resolution varies from 70.5\,arcmin for the lowest frequency band to 17.9\,arcmin FWHM for the highest frequency band. The \LB\ simulations assume tophat bandpasses with relative widths of 0.23 or 0.30 for each band. The center frequency, bandpass width, beam size, total number of bolometers, noise equivalent temperature (NET) for the total detector array, polarisation sensitivity, and the telescope on which each frequency band is located can be found in \cref{tab:LiteBIRD}. The polarisation sensitivity takes into account an observation efficiency of $\eta=0.77$ caused by an expected data loss from the observation duty cycle, cosmic rays, and an added margin. 

The whole satellite itself will be spinning at 0.05 rotations per minute (rpm) at a spin angle $\beta=50\deg$ and a precession angle $\alpha=45\deg$  \citep{Sugai_2020,Takase_2024}. The three HWPs will spin at 46, 39 and 61 rpm, respectively. The sampling rate for all three telescopes are 19\,Hz. This leaves us with $8\times 10^{12}$ detector samples or almost 20\,TB of uncompressed raw data in total for the sky signal, not including metadata \citep{LB_PTEP}.

As of today, \LB\ is the only funded next-generation full-sky CMB experiment and will observe $B$ modes over the multipoles $2\leq \ell \leq 200$. Therefore one of its advantages over ground based experiments is that it will be able to detect both the reionisation bump below multipoles of $\ell \simeq10$ and the recombination peak at $\ell \simeq 80$, while ground based experiments are limited to only the recombination peak.

\begin{table*}[t] 
  \begingroup
  \newdimen\tblskip \tblskip=5pt
  \caption{
    \LB\ parameters for instrument model V1 from July 2020 \citep{LB_PTEP}. 
    The bolometers are divided between two instruments where two numbers are listed. 
  }
  \label{tab:LiteBIRD}
  \nointerlineskip
  \vskip -3mm
  \footnotesize
  \setbox\tablebox=\vbox{
    \newdimen\digitwidth
    \setbox0=\hbox{\rm 0}
    \digitwidth=\wd0
    \catcode`*=\active
    \def*{\kern\digitwidth}
    \newdimen\signwidth
    \setbox0=\hbox{-}
    \signwidth=\wd0
    \catcode`!=\active
    \def!{\kern\signwidth}
 \halign{
      \hbox to 7.5cm{#\leaderfil}\tabskip 1em&
      \hfil#\hfil\tabskip 1em&
      \hfil#\hfil\tabskip 1em&
      \hfil#\hfil\tabskip 1em&
      \hfil#\hfil\tabskip 1em&
      \hfil#\hfil\tabskip 1em&
      \hfil#\hfil\tabskip 1em&
      \hfil#\hfil\tabskip 0pt\cr
    \noalign{\doubleline}
      \omit\hfil\hskip 10pt\textsc{Center frequency}\hfil&
      \omit\hfil\textsc{Bandpass $\delta\nu$}\hfil&        
      \omit\hfil\textsc{Beam size}\hfil&
      \omit\hfil\textsc{Total number of}\hfil&
      \omit\hfil\textsc{NET array}\hfil&
      \omit\hfil\textsc{Polarisation Sensitivity}\hfil&
      \omit\hfil\textsc{Telescope}\hfil\cr
      \omit\hfil\hskip 10pt\textsc{[\,GHz ]\,}\hfil&
      \omit\hfil\textsc{[\,GHz ]\,}\hfil&
      \omit\hfil\textsc{[\,arcmin ]\,}\hfil&
      \omit\hfil\textsc{Bolometers}\hfil&
      \omit\hfil\textsc{[\,$\mu$K\ rts ]\,}\hfil&
      \omit\hfil\textsc{[\,$\mu$K\ arcmin ]\,}\hfil&
      \omit\hfil\textsc{}\hfil\cr
      \noalign{\vskip 4pt\hrule\vskip 4pt}
      \noalign{\vskip 2pt}
      \omit\hskip 38pt 40	& 12  & 70.5		    & 48		  & 18.50		   & 37.42	& LFT \cr
      \omit\hskip 38pt 50 	& 15  & 58.5		    & 24		  & 16.54		   & 33.46	& LFT \cr
      \omit\hskip 38pt 60   & 14  & 51.1		    & 48		  & 10.54		   & 21.32	& LFT \cr
      \omit\hskip 38pt 68   & 16  & (41.6, 47.1)	& (144, 24)	  & (9.84, 15.70)  & 16.87	& LFT \cr
      \omit\hskip 38pt 78   & 18  & (36.9, 43.8)	& (144, 48)	  & (7.69, 9.46)   & 12.07	& LFT \cr
      \omit\hskip 38pt 89	& 20  & (33.0, 41.5)	& (144, 24)   & (6.07, 14.22)  & 11.30	& LFT \cr
      \omit\hskip 35pt 100	& 23  & (30.2, 37.8)	& (144, 366)  & (5.11, 4.19)   & 6.56	& L/MFT \cr
      \omit\hskip 35pt 119	& 36  & (26.3, 33.6)	& (144, 488)  & (3.8, 2.82)	   & 4.58	& L/MFT \cr
      \omit\hskip 35pt 140	& 42  & (23.7, 30.8)	& (144, 366)  & (3.58, 3.16)   & 4.79	& L/MFT \cr
      \omit\hskip 35pt 166 	& 50  & 28.9		    & 488         & 2.75		   & 5.57	& MFT \cr
      \omit\hskip 35pt 195	& 59  & (28.0, 28.6)	& (366, 254)  & (3.48, 5.19)   & 5.85	& M/HFT \cr
      \omit\hskip 35pt 235	& 71  & 24.7		    & 254		  & 5.34		   & 10.79	& HFT \cr
      \omit\hskip 35pt 280	& 84  & 22.5		    & 254		  & 6.82		   & 13.80	& HFT \cr
      \omit\hskip 35pt 337	& 101 & 20.9		    & 254		  & 10.85		   & 21.95	& HFT \cr
      \omit\hskip 35pt 402	& 92  & 17.9		    & 338		  & 23.45		   & 47.45	& HFT \cr
      \noalign{\vskip 4pt\hrule\vskip 2pt}
      } }
  \endPlancktablewide \endgroup
\end{table*}

%% file: Algorithm.tex
\section{Algorithm}
\label{sec:Algorithm}

The code used to analyse the simulated \LB\ data in this work is called \commanderthree, and it uses an end-to-end Bayesian approach for the CMB analysis. Starting with TOD we do calibration and map-making, and component separation iteratively to sample the full parameter space. Only a brief overview of the \commanderthree\ algorithm will be presented here. For a full description, see refs.~\citep{bp01}, \citep{bp03} and \citep{watts2023cosmoglobe}.

\subsection{Data model}
Following the usual parametric Bayesian analysis approach, we start by writing an explicit data model, which for this paper is given for a detector $j$ and time sample $t$ as
\begin{equation}
    d_{t,j} = g_{t,j}P_{tp,j}\left[B^{\mathrm{symm}}_{pp',j}\sum_{c} M_{cj}(\beta_{p'})\a^c_{p'}  + s^{\mathrm{orb}}_{j}\right] 
    	+ \n^{\mathrm{corr}}_{t,j} + \n^{\mathrm{w}}_{t,j}.
  \label{eq:todmodel_LB}
\end{equation}
Here $d_{t,j}$ is the measured data per time sample per detector. The current simulations assume ideal gain, giving us $g_{t,j}=1$. The pointing matrix $P_{tp,j}$ contains $(1,\cos{2\psi}, \sin{2\psi})$ for the column corresponding to the Stokes parameters $\{T,Q,U\}$ for pixel $p$ and row corresponding to time $t$, and is zero elsewhere, where $\psi$ is the polarisation angle, $T$ is the intensity of the signal and $\{Q,U\}$ represent the polarised signal. The beam convolution matrix, $B_j$, contains both the beam and the pixel window function caused by our maps being pixelised with the HEALPix projection \citep{gorski2005}. The simulations use perfect Gaussian beams, with no beam asymmetries. The mixing matrix, $M_{cj}(\beta_{p'})$, is dependent on the spectral parameters $\beta$, for component $c$ and detector $j$ relative to some reference frequency. This matrix is diagonal in pixel space. The amplitudes, $\a^c_{p'}$, for sky component $c$ in pixel $p$ is measured in brightness temperature units, and the orbital CMB dipole, $s^{\mathrm{orb}}$, is  measured in $\mathrm{K_{cmb}}$. All bandpasses are assumed to be perfectly known. The noise is divided into correlated noise, $\n^{\mathrm{corr}}$, and white noise, $\n^{\mathrm{w}}$. The white noise level is caused mainly by thermal noise, and it is thus heavily temperature dependent with lower detector temperatures leading to a lower white noise level. It also decreases with the number of observations $N_{\mathrm{obs}}$ of pixel $p$, as $\n^{\mathrm{w}}_p \propto 1/\sqrt{N_{\mathrm{obs}}}$. 

The astrophysical sky model adopted in this paper consists of CMB and foreground emission from our Galaxy. The relevant Galactic foregrounds for this analysis are synchrotron radiation, thermal dust emission and free-free emission. Both synchrotron and thermal dust are polarised while free-free is not, so free-free is only included in intensity.

We assume that the CMB ($\a_{\mathrm{CMB}}$) spectral energy density (SED) can be approximated as a blackbody. Synchrotron radiation dominates the polarised signal at the lowest \LB\ frequencies, and is described by a power law model with a free amplitude per pixel, $\a_{\mathrm{s}}$. Similarly, thermal dust emission dominates the highest frequencies, and is modelled with a modified blackbody spectral energy density multiplied with a free amplitude per pixel, $\a_{\mathrm{d}}$ . Free-free radiation ($\a_{\mathrm{ff}}$) is based on the \Planck-2015 model \citep{planck2014-a12}. In total, our sky model SED in brightness temperature is described by
\begin{align}
	\s_{\mathrm{RJ}} = & \a_{\mathrm{CMB}} \frac{x^2e^x}{(e^x-1)^2}  \nonumber \\ 
	+ &   \a_{\mathrm{s}}\left(\frac{\nu}{\nu_{0,\mathrm{s}}} \right)^{\beta_{\mathrm{s}}} \nonumber\\
	+ & \a_{\mathrm{d}} \left( \frac{\nu}{\nu_{0,\mathrm{d}}} \right)^{\beta_{\mathrm{d}}+1}
	\frac{e^{h\nu_{0,\mathrm{d}}/kT_{\mathrm{d}}}-1}{e^{h\nu/kT_{\mathrm{d}}}-1} \label{eq:SED}\\
    + & \a_{\mathrm{ff}}\left(\frac{\nu_{0,\mathrm{ff}}}{\nu}\right)^2 \frac{g_{\mathrm{ff}}(\nu;T_{\mathrm{e}})}{g_{\mathrm{ff}}(\nu_{0,\mathrm{ff}};T_{\mathrm{e}})} \nonumber \\
    + & m_{\nu},\nonumber
\end{align}
where $\a_i$ are the amplitudes of the signal components at frequency $\nu$, and $\nu_{0,i}$ are the reference frequencies for each component; $x=h\nu/kT_0$, where $h$ and $k$ are the Planck and Boltzmann constants respectively and $T_0 = 2.7255$\,K is the CMB temperature \citep{fixsen2009};
$\beta_i$ are the spectral indices for synchrotron and thermal dust and  $T_{\mathrm{d}}$ is the thermal dust temperature; $g_{\mathrm{ff}}$ is the so-called Gaunt factor for the free-free emission described by ref.~\citep{dickinson2003} dependent on the electron temperature $T_{\mathrm{e}}$; and $m_{\nu}$ is the CMB monopole. Explicitly, this model defines the sum over components in eq.~(\ref{eq:todmodel_LB}) with $\beta\equiv \{\beta_{\mathrm{s}}, \beta_{\mathrm{d}}, T_{\mathrm{d}}, T_{\mathrm{e}}\}$.

The lowest frequency, 40\,GHz, is chosen as the reference frequency for synchrotron, the highest frequency, 402\,GHz, is set as the thermal dust reference frequency, and 50\,GHz is chosen as reference frequency for free-free. Fitting spectral indices will be a minor step in terms of computational costs compared to the TOD sampling, as \LB\ has a relative low spatial resolution and a large data volume and will thus be highly TOD dominated in terms of analysis costs, see \cref{ss:runningcost} for more details. The spectral indices for synchrotron and dust as well as the dust temperature are therefore kept at their true input values for simplicity.

In total, our parameter space consists of $\omega =\{\a_{\mathrm{CMB}}, \a_{\mathrm{s}}, \a_{\mathrm{d}}, \a_{\mathrm{ff}}, m_{\nu}, \n_{\mathrm{corr}}\}$ where $\a_{\mathrm{ff}}$ is unpolarised. We use a TOD processing mask with $f_\mathrm{sky}=85\%$ generated using a sky cut-off after smoothing a 337\,GHz band residual map from a previous iteration (that contained the unmodelled residual signal) with a $5\deg$ beam. 
We denote the set of all free parameters in this data model by $\omega$.

We note that the sky model defined by eq.~(\ref{eq:SED}) defines the parametric model that is actually fitted to the data, and this is obviously much simpler than the real sky (e.g.,~\citep{meisner2015,Green_2019,Liu_2025,Vacher_2025}), and, indeed, even than the simulations analysed in this paper; see \cref{ss:Foreground}. Choosing the optimal fitting model is typically an iterative process informed by the $\chi^2$ as a function of sky position, and involves a trade-off between minimising residuals and model degeneracies. For instance, the current baseline \LB\ configuration has 15 main frequencies between 40 and 402\,GHz, and it can therefore at most allow 14 free parameters without becoming fully degenerate. In practice, fewer parameters can be fitted robustly. On the other hand, joint analysis with external data can in principle increase this indefinitely.

\subsection{Bayesian analysis and Gibbs sampling}
\label{ss:gibbs}
\commanderthree\ is built on a Bayesian framework, with the goal of estimating the posterior distribution $P(\omega\mid \d)$  using Bayes' theorem, 
\begin{equation}
  P(\omega\mid \d) = \frac{P(\d\mid \omega)P(\omega)}{P(\d)} \propto
  \mathcal{L}(\omega)P(\omega),
  \label{eq:jointpost}
\end{equation}
where $P(\d\mid \omega)\equiv\mathcal{L}(\omega)$ is the likelihood, $P(\omega)$ is the prior for all free parameters $\omega$, and $P(\d)$ is called the evidence. The latter acts simply as an overall normalisation constant, since it is independent of $\omega$, and it is therefore omitted in the following. 

Since the number of free parameters in $\omega$ is very large (on the order of millions), we resort to Markov Chain Monte Carlo~(MCMC) methods to map out $P(\omega \mid \d)$.
In particular, Gibbs sampling \citep{geman:1984,Gelman03} serves as the main computational engine, while other methods such as Metropolis and Metropolis-Hastings sampling are also used for individual sampling steps. In Gibbs sampling, we sample at any given time only one parameter or a subset of parameters while the rest are kept fixed. To explore the full parameter space this way, our sampling is executed iteratively. The Gibbs sampling chain in \commanderthree\ for this work is as follows:
\begin{align}
\n_{\mathrm{corr}} &\leftarrow P(\n_{\mathrm{corr}}	\mid \d,\a  )\\
\a 				   &\leftarrow P(\a \mid\d, \n_{\mathrm{corr}}	   ),
\end{align}
where the arrows means that the value for the sampled parameter is drawn from the distribution to the right. Both of these steps are performed with Gaussian multivariate samplers, as described by refs.~\citep{bp06} and \citep{bp13}, respectively. 

Finally, we note that a future full \LB\ analysis will obviously include many more parameters than those considered here, and each of those will require a conditional separate sampling step in the above Gibbs chain. However, experience from previous \BP\ and Cosmoglobe analyses shows that the correlated noise sampler (which performs the same function as a maximum likelihood mapmaker in a traditional pipeline) and the sky signal amplitude sampler (which is the core of the component separation procedure) dominate the overall runtime. This will be even more true for \LB, for which the TOD volume will be much higher than for current experiments, though the resolution is actually lower. We do not anticipate new sampling steps will be more expensive than the existing ones, nor do we expect a big effect on our burn-in time from these additions. For the purposes of this paper, which focuses on computational feasibility, the two steps discussed above are therefore by far the most essential.

%% file: Simulations.tex
\section{Simulations}
\label{sec:sims}
We analyse the first generation time-ordered data produced with the dedicated \LB\ Simulation Framework (LBS)\footnote{https://github.com/litebird/litebird\_sim} \citep{tomasi:2025}. The main instrumental properties of the \LB\ instruments are summarised in \cref{tab:LiteBIRD}. Based on these, maps of CMB and foregrounds have been co-added, tophat bandpass-integrated and beam convolved for the instrumental beam for each frequency band to represent our microwave sky. The LBS is then used to generate TOD simulations for each detector given \LB's scanning strategy and the specifics of each detector. We will now go into more detail regarding the sky maps for Galactic foregrounds in \cref{ss:Foreground} and extra Galactic sources in \cref{ss:exgal}, before we present the simulation pipeline and the specifics of our simulations.

\subsection{Galactic foregrounds}
\label{ss:Foreground}
The simulated sky maps have been generated using the map-based simulation module (MBS) provided by LBS, which acts as a wrapper around the Python Sky Model (PySM)\footnote{https://github.com/galsci/pysm} package \citep{Thorne_2017, Zonca_2021}. Specifically, the model considered accounts for realistic Galactic  foreground emission maps using the PySM \texttt{d1s1a1f1co1} model. A thermal dust component (\texttt{d1}) is modelled as a single-component modified blackbody based on the maps from the \Planck\ 2015 analysis \citep{planck2014-a12} with varying temperature and spectral index across the sky. A synchrotron  component (\texttt{s1}) is modeled as a power law also with a varying spectral index but no curvature, based on Haslam 408\,MHz and WMAP data \citep{delabrouille_2013, miville_2008}. An unpolarised Anomalous Microwave Emission (AME) component (\texttt{a1}) is included as two spinning dust populations from the \citep{planck2014-a12} model. Unpolarised free-free emission (\texttt{f1}) is also modeled based on ref.~\citep{planck2014-a12} in terms of spatial morphology, but employing a constant spectral index of $\beta_{\mathrm{ff}}=-2.14$ rather than the full non-linear model. Finally,  a model of  Galactic CO emission (\texttt{co1}) is modeled involving the first three CO rotational lines, i.e.  $J:1-0, 2-1, 3-2$,  whose center frequencies are respectively at \{$115.3$, $230.5$, $345.8$\}\,GHz  \citep{puglisi_2017, planck2013-p03a}.

\subsection{Extra Galactic sources}
\label{ss:exgal}
The modelling of extra Galactic emission has been recently included in the PySM models. In the \LB\ simulations we have employed consistently the models coming from the WebSky simulations \citep{Stein_2020}, where the emission is obtained by populating dark matter haloes with halo-occupation distribution consistent with the observations. Starting from the dark matter distributions, ref.~\citep{Stein_2020} produced  cosmic infrared background (CIB), thermal (tSZ) and kinetic Sunyaev–Zeldovich emission (kSZ),  and a lensing convergence map. 

Radio sources have recently been included by ref.~\citep{Li_2022}, with fluxes (both in total intensity and polarisation) to be consistent with the latest  observations (see ref.~\cite[Fig.~5]{Li_2022} and ref.~\cite{puglisi_2018}) and latest number counts \citep{lagache_2020}. 
Finally, we synthesise a  CMB map from the \Planck\ 2018 cosmological parameters, and lens it with the WebSky convergence map, so that the lensing component correlates with the other large-scale structure probes.

\subsection{\LB\ Instrument Model and TOD simulation pipeline}
\label{ss:IMO}
The simulations used in the following have been produced using the LBS, and the current paper represents the first non-LBS application of these TOD simulations. Indeed, an important motivation for the current paper is simply to confirm that these simulations can be readily used by external codes.

The LBS interfaces with the Instrument Model database, the IMo\footnote{https://github.com/litebird/litebird\_imo (accessible by \LB\ members upon request)}, that contains information about the \LB\ satellite, such as parameters for the scanning strategy, HWP rotation speed, detector noises and instrumental beams. The first step of the simulation pipeline is to load the IMo information. With this, the pipeline computes the quaternions to pass from the ecliptic to the detectors reference frames. The quaternions and the HWP rotation speed are finally used to calculate the pointing information (colatitude and longitude) and the polarisation angle for each detector considered in the simulation for the whole duration of the simulation. The following TODs are then produced and saved to disk:
\begin{itemize}
    \item \texttt{tod\_cmb}, containing the CMB signal obtained by scanning the input CMB maps using the pointing information,
    \item \texttt{tod\_fg}, containing the foregrounds signal obtained by scanning the input foregrounds maps using the pointing information,
    \item \texttt{tod\_dip}, containing the CMB dipole signal, (the dipole model used, following the \lbs\ definition, is \texttt{TOTAL\_FROM\_LIN\_T}\footnote{https://litebird-sim.readthedocs.io/en/master/dipole.html}),
    \item \texttt{tod\_wn}, containing white noise obtained using the detector noise information,
    \item \texttt{tod\_wn\_1f\_30mHz}, containing correlated $1/f$ and white noise obtained using the detector noise information and the realistic case with a knee frequency $f_{knee}$ of 30 mHz,
    \item \texttt{tod\_wn\_1f\_100mHz}, same as \texttt{tod\_wn\_1f\_30mHz}, but with the pessimistic  knee frequency $f_{knee}$ of 100 mHz.
\end{itemize}

All the above components are produced individually, and later co-added for a full TOD. Due to memory and CPU hours budget limitations, not all the \LB's detectors have been considered, nor the entire mission time (3 years). Specifically, for frequency channels with more than 48 detectors, only about one third of all bolometers were included, choosing a configuration that preserves the focal plane symmetry; for channels that have fewer than 48 detectors, all bolometers were simulated. Furthermore, only one third of the mission time (1 year) has been simulated. 

\subsection{Final data selection and compression}
\label{ss:LB_sims}
The expected amount of data from \LB\ far exceeds that of any previous CMB satellite mission. For instance, while \Planck\ LFI had 22 radiometers observing for 4 years with a sampling rate between 32 and 79\,Hz \citep{planck2016-l01}, \LB\ will have a total number of 4508 bolometers observing for three years with a sampling rate of 19\,Hz. To comfortably fit the simulations on our nodes with 1.5\,TB of memory during the Bayesian Commander-based analysis, we adopted 1\,TB as the maximum data volume for our simulated \LB\ data, and included only a subset of the full simulated data volume, comprising 4 detectors at each distinct channel.  The noise level per detector has been scaled down to account for both the decreased number of detectors and the reduced observation time, so that the total sensitivity per frequency band will be comparable to the full experiment. 

\begin{figure}[t]
    \centering
    \includegraphics[width=\linewidth]{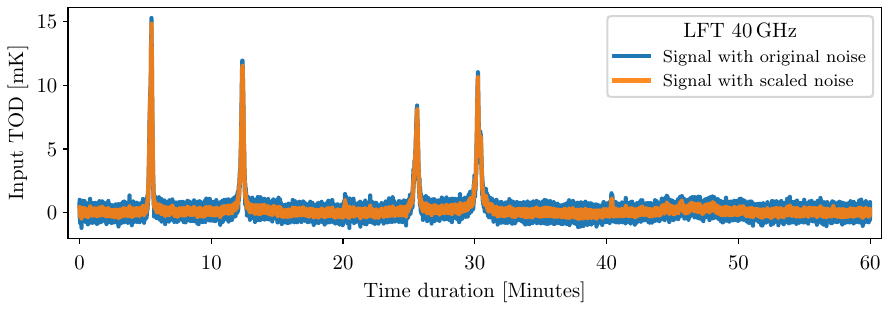} 
  \caption{Time ordered data for 1 hour of samples for the 40\,GHz channel. The plot includes the total signal (CMB, noise and the foreground emissions) for the simulated TOD (blue). Also plotted is the total signal with the scaled down noise levels that compensate for reducing the number of detectors from 48 to 4 (orange). Units are in millikelvin.}
  \label{fig:TOD_reduced_noise}  
\end{figure}

\begin{figure} 
    \includegraphics[width=\linewidth]{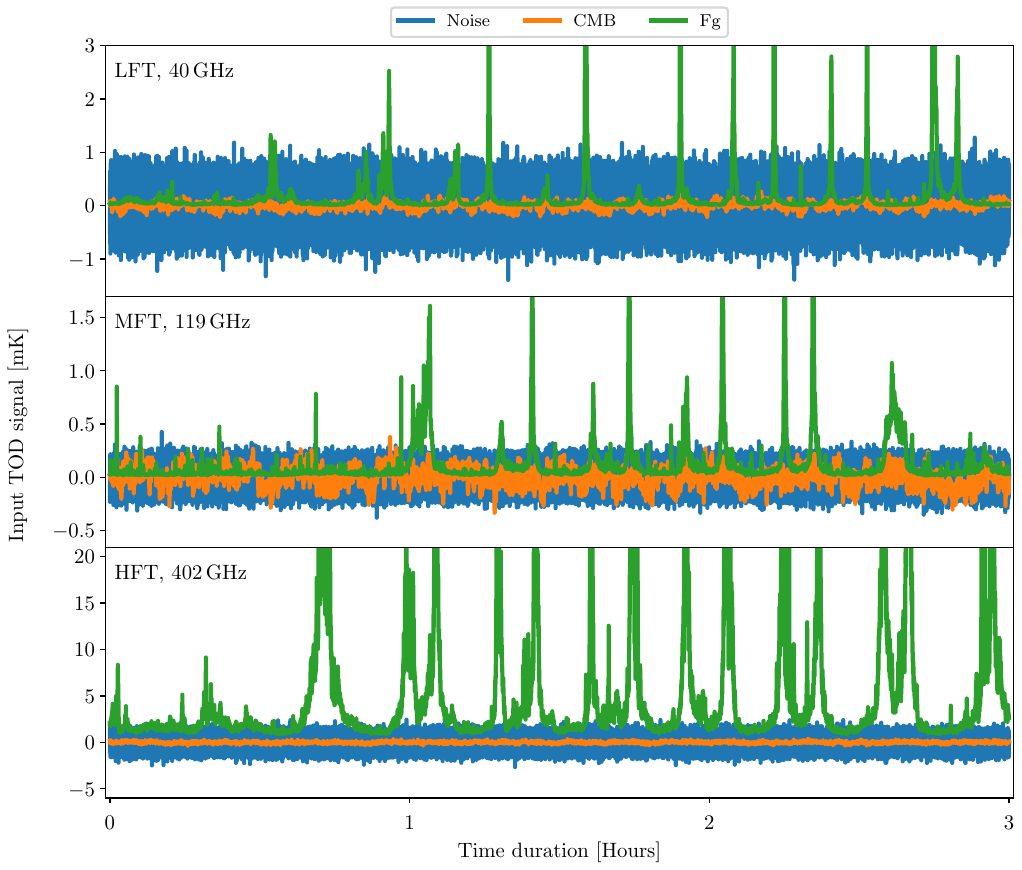}
  \caption{Time ordered data for the same three hours worth of samples for one detector for the LFT 40\,GHz channel (top), the MFT 119\,GHz channel (middle) and the HFT 402\,GHz channel (bottom). The plots include the CMB (orange), white and correlated noise (blue) and the foregrounds (green). The peaks correspond to scanning the Galaxy plane, where synchrotron dominates the lowest frequency, the middle frequency has the lowest level of foregrounds, and the highest frequency channel foregrounds are dominated by thermal dust.}
  \label{fig:sample_timeline}  
\end{figure} 

Our analysis includes a total of 88~detectors and one year of observations. These data are stored as Hierarchical Data Format (HDF) files including pointing, flags, TOD values and polarisation per sample, as well as common meta data describing the instrument. As described by ref.~\citep{bp03}, we compress these data to minimise the memory footprint, noting that all data are stored in RAM during processing. The total data volume for the selected data set is 1.55\,TB without compression and 470\,GB with compression (see \cref{tab:datavolume}), compression, thus, saves a factor of 3.4 in terms of RAM requirements. In addition, we note that the current \lbs\ TOD do not yet include an analog-to-digital conversion~(ADC) step, and the resulting samples are therefore currently stored as floating point numbers rather than integers. Integers are obviously much more compressible than floats, and once a proper ADC procedure has been implemented, the compression efficiency will increase further.

Scaling the current data to the full experiment with 4508 detectors and 3 years of data results in a total data volume of 238\,TB without and 70\,TB with compression. The scaling should, however, be slightly better than linear. Housekeeping data, for example, does not scale with the number of detectors, and the compression implemented in this work was originally optimised for the \Planck\ LFI data. Other compression alternatives are currently being considered for \LB, and 70\,TB is thus a strict upper estimate for the final compressed data volume.

%% file: Results.tex
\section{Results}
\label{sec:Results}

We now present the results from the analysis on the simulations described in \cref{sec:sims} using the method described in \cref{sec:Algorithm}. 
Note that there are more foreground components present in the simulations than we currently fit for using \commanderthree. In order to keep the analysis straightforward, only the most dominant foregrounds have been included. Doing a complete analysis with more complicated foreground components, and instrumental and systematic effects is left for future work.

During the course of the following analysis, we have produced several hundreds Gibbs samples involving all \LB\ channels. However, in the following we show results from one frequency channel for each telescope, where we have chosen the lowest (40\,GHz) and the highest frequency (402\,GHz) for LFT and HFT respectively, and for MFT we have chosen the frequency with the lowest total noise values, the 119\,GHz channel. Intermediate channels look qualitatively similar, and different Gibbs samples are indistinguishable up to instrumental noise. We first inspect the TODs before we show the resulting frequency maps and finally the data volume and the running costs needed to do the analysis. 

\begin{figure}[t]
    \centering
    \includegraphics[width=\linewidth]{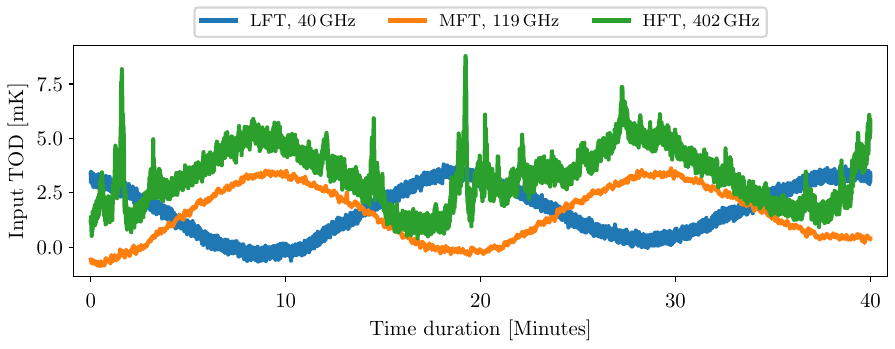} 
  \caption{Time ordered data for 40\,min for the 40\,GHz channel (blue), the MFT~119\,GHz channel (orange), and the 402\,GHz channel (green) including CMB, foregrounds, white and correlated noise, and the dipole. The dipole signal causes the large sinusoidal pattern and the other components are responsible for the thickness of the lines. The spikes of the 402\,GHz channel are foreground signals. The dipole in the LFT channel has the opposite sign compared to the MFT and HFT, which is caused by the telescope pointing in nearly opposite sky direction.}
  \label{fig:TOD_dipole}  
\end{figure}

\subsection{Time-ordered data}
\label{ss:scanning}

\Cref{fig:TOD_reduced_noise} shows the total TOD including noise, foreground emission and the CMB signal for one hour of data for one detector in the LFT~40\,GHz channel. The blue line shows the total signal of a single detector from the simulation containing the full set of detectors.
The expected noise level from real data will have a $\sqrt{3}$ stronger noise signal due to the noise reduction we applied when scaling the mission duration from three to one year. The orange line shows the same TOD segment but with the reduced noise level as described in \cref{ss:LB_sims}. 

\begin{table*} [t]
  \begingroup
  \newdimen\tblskip \tblskip=5pt
  \caption{
  	Computational costs for 88 detectors and 1 year (this work) in the left column and estimations for the full mission with 4508 detectors and 3 years in the right column. The costs are split into RAM requirements and processing costs in terms of CPU hours. The processing costs are split into costs per run for initialisation steps and costs per sample for TOD processing and for component separation. The numbers are based on running the analysis on a 2.6\,GHz AMD cluster node with 128 cores and 2\,TB of memory. 
  }
  \label{tab:datavolume}
  \nointerlineskip
  \footnotesize
  \setbox\tablebox=\vbox{
    \newdimen\digitwidth
    \setbox0=\hbox{\rm 0}
    \digitwidth=\wd0
    \catcode`*=\active
    \def*{\kern\digitwidth}
    \newdimen\signwidth
    \setbox0=\hbox{-}
    \signwidth=\wd0
    \catcode`!=\active
    \def!{\kern\signwidth}
 \halign{
      \hbox to 6.5cm{#\leaderfil}\tabskip 1em&
      \hfil#\tabskip 1em&
      \hfil#\tabskip 1em&
      \hfil#\tabskip 1em&
      \hfil#\tabskip 1em&
      \hfil#\tabskip 0pt\cr
    \noalign{\doubleline}
      \omit\textsc{}\hfil&
      \omit\hfil\textsc{88 detectors}\hfil&
      \omit\hfil\textsc{4508 detectors}\hfil\cr
      \omit\textsc{}\hfil&
      \omit\hfil\textsc{1 year}&
      \omit\hfil\textsc{3 years}\cr  
      \noalign{\vskip 4pt\hrule\vskip 4pt}
      \noalign{\vskip 2pt}
      \multispan3\hskip 5pt\textit{Data volume}\hfil\cr
      \hskip 15pt Uncompressed TOD volume       & 1.55 TB   & 238 TB\cr
      \hskip 15pt Compressed TOD volume         & 470 GB    & 70 TB\cr
      \hskip 15pt Non-TOD-related RAM usage     & 230 GB    & 230 GB\cr
      \hskip 15pt {\bf Total RAM requirements}  & {\bf 700 GB} & {\bf 70 TB}\cr      
      \noalign{\vskip 2pt}      
      \multispan3\hskip 5pt\textit{Processing time (cost per run)}\hfil\cr
      \noalign{\vskip 2pt}
      \hskip 15pt TOD initialisation            & 9.5 h    & 1500 h\cr
      \hskip 15pt Other initialisation          & 6.7 h   & 7 h\cr
      \hskip 15pt {\bf Total initialisation}    & {\bf 16.3 h} & {\bf 1500 h}\cr
      \noalign{\vskip 2pt}      
      \multispan3\hskip 5pt\textit{Gibbs sampling steps (cost per sample)}\hfil\cr
      \hskip 15pt Huffman decompression         & 0.6 h   & 90 h\cr
      \hskip 15pt TOD projection ($P$ operation)& 1.2 h  & 200 h\cr
      \hskip 15pt Orbital dipole ($\s_{\mathrm{orb}}$)& 0.5 h & 80 h\cr
      \hskip 15pt Correlated noise sampling ($\n_{\mathrm{corr}}$)& 6.1 h & 1000 h\cr
      \hskip 15pt TOD binning ($P^t$ operation)& 0.1 h   & 10 h\cr
      \hskip 15pt Sum of other TOD processing   & 9.8 h    & 1500 h\cr
      \hskip 15pt {\bf TOD processing cost per sample} & {\bf18.3 h} & {\bf 3000 h}\cr
      \noalign{\vskip 2pt}
      \hskip 15pt Amplitude sampling, $P(\a\mid \d, \omega\setminus\a)$  & 5.9 h & 10 h\cr
      \hskip 15pt Spectral index sampling, $P(\beta\mid \d, \omega\setminus\beta)$  & &20 h\cr
      \hskip 15pt Other steps                   & 1.1 h   & 1 h\cr
      \noalign{\vskip 2pt}
      \hskip 15pt {\bf Total cost per sample}   & {\bf25.4 h}  & {\bf 3000 h}\cr
      \noalign{\vskip 4pt\hrule\vskip 2pt}
      } } 
  \endPlancktablewide \endgroup
\end{table*} 

Three hours of simulated TOD for one detector each in three different frequency channels can be seen in \cref{fig:sample_timeline}; LFT~40\,GHz in the top panel, MFT~119\,GHz in the middle panel and HFT~402\,GHz in the bottom panel, all for the same three hours of data. The plots show the scaled down noise including both white and correlated noise (blue), the total foreground signals (green) and the CMB signal (orange) individually. The first thing to be noticed are the green peaks corresponding to the scanning of the Galactic plane which typically happens several times each hour. The noise levels in all channels have been scaled down to preserve the noise level in the resultant maps while compensating for the reduced mission duration and number of channels in the simulations, and thus the noise shown in these three plots will not be representative of what we expect for the real data. It will not be comparable between the frequency channels either since the 119\,GHz channel with its 488 detectors and the 402\,GHz channel with its 338 detectors will have their detector noise level scaled down by approximately a factor of 19 and 16, respectively, while the 40\,GHz channel with its 48 detectors will scale the noise down by a factor of 6. The CMB and foreground signals are however representative for how they will be observed by each of the three frequency bands. 

The signal amplitudes in \cref{fig:sample_timeline} are all given in millikelvin, but the strength of the signal varies for each frequency. The thermal dust dominated 402\,GHz channel has a 70 times stronger foreground signal than the 119\,GHz channel with the weakest foreground signal, while the synchrotron dominated 40\,GHz has a 3.5 times stronger foreground signal.

\Cref{fig:sample_timeline} also shows the MFT and HFT channels scanning the Galactic plane simultaneously as these two telescopes point in the same sky direction, while the foreground peaks in the LFT channel are as expected located at different times due to the LFT channel pointing in a direction that is $100 \deg$ separated from the HFT~\citep{LB_PTEP, Takase_2024}. This can also be observed in \cref{fig:TOD_dipole} where the dipole is added to the total TOD signal and the dipole varies with the opposite sign for the LFT channel as for the frequency channels on the other two telescopes.

\subsection{Frequency maps}
\label{ss:maps}

\begin{figure} [t]
 \center
    \includegraphics[width=0.32\linewidth]{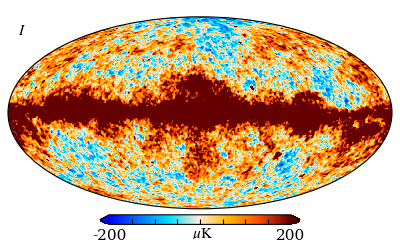}
    \includegraphics[width=0.32\linewidth]{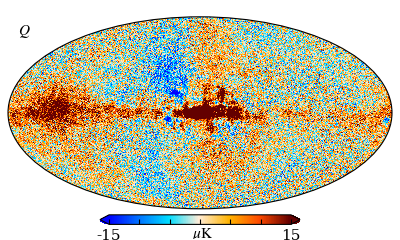}
    \includegraphics[width=0.32\linewidth]{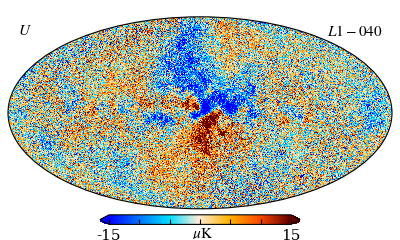}\\
    \includegraphics[width=0.32\linewidth]{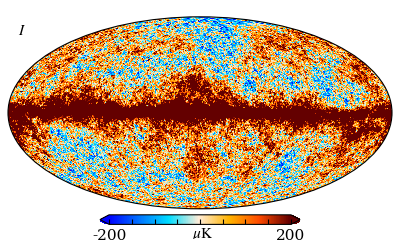}
    \includegraphics[width=0.32\linewidth]{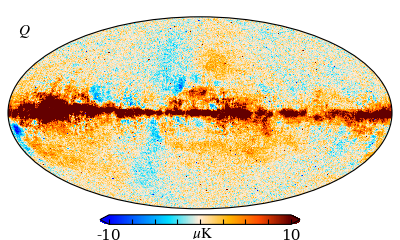}
    \includegraphics[width=0.32\linewidth]{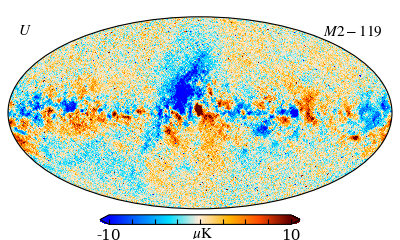}\\
    \includegraphics[width=0.32\linewidth]{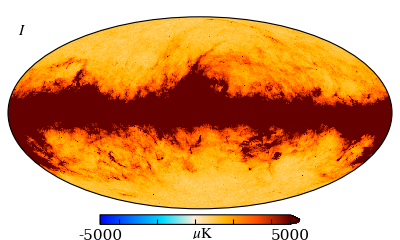}
    \includegraphics[width=0.32\linewidth]{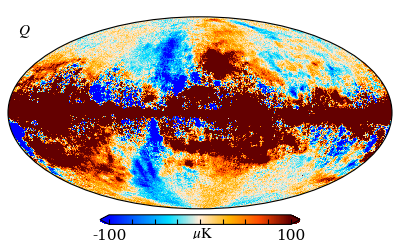}
    \includegraphics[width=0.32\linewidth]{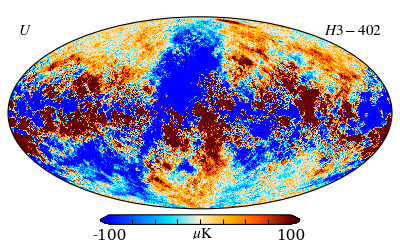}\\
  \caption{Frequency maps of a single Gibbs sample for three frequency bands for 40\,GHz (top), 119\,GHz (middle) and 402\,GHz (bottom), represented by Stokes $I$ (left column), $Q$ (middle) and $U$ (right column) parameters. The lowest frequency can be seen to be dominated by synchrotron while the highest one is dominated by thermal dust. The middle frequency has the lowest sky signal. All units are in microkelvin.}
  \label{fig:freq_maps}  
\end{figure}

\Cref{fig:freq_maps} shows sky maps for three frequency bands, with the temperature maps (Stokes $I$) in the left column and polarisation (Stokes $Q$ and $U$) in the middle and right columns. The LFT~40\,GHz is in the top row, MFT~119\,GHz in the middle row and HFT~402\,GHz frequency channel in the bottom row. All maps are plotted at $N_{\mathrm{side}}=512$ in Galactic coordinates. 

The polarised 40\,GHz maps can be seen dominated by synchrotron and the 402\,GHz maps are dominated by thermal dust, while the 119\,GHz maps have a mix of both foreground signals. We clearly see the Galactic plane and we can also see the white noise in the 40\,GHz maps as the sky signal is weaker than the 402\,GHz sky signal and the polarisation sensitivity is significantly lower than for the 119\,GHz channel, see \cref{tab:LiteBIRD}. The strength of the foreground signals also varies. Note for instance that the polarised 402\,GHz signal is plotted with a 10 times higher temperature range than for the 119\,GHz channel, in accordance with the TOD plots in \cref{fig:sample_timeline}.

\subsection{Computational costs}
\label{ss:runningcost}
Next, we consider the main result from this paper, namely the computational costs for running \commanderthree\ on the simulated \LB\ TOD. These are listed in \cref{tab:datavolume} for the main operations in units of CPU hours. Total computational costs are divided into one-time processing costs for a single run (including initialisation steps) and costs per Gibbs sample. The cost of each Gibbs sampling step has been further divided into TOD processing step and parameter sampling (component separation) step. Adding more data in time domain will not affect the cost of the component separation step as the component separation happens in map space, and the number of maps will stay fixed as long as we only work with the \LB\ data set. 

In the current run, all maps are pixelised with a HEALPix \citep{gorski2005} resolution parameter of $N_{\mathrm{side}}=512$ (corresponding to $7\arcm$ pixels), but we anticipate that $N_{\mathrm{side}}=1024$ will be used for the highest frequency channels in a final production run. Accounting for this, we estimate that the foreground amplitude sampling step for the full \LB\ data set will increase from 6 to 10\,CPU~hours per Gibbs sample.  

With our reduced data set the TOD processing costs are of the same order of magnitude as the component separation costs, as shown in \cref{tab:datavolume}. Scaling up the amount of TOD to the full \LB\ experiment, however, makes the component separation part of the analysis strongly subdominant. The most expensive single Gibbs sample step in terms of computational cost is the correlated noise sampling with 6\,CPU~hours or 24\%\ of the total CPU hours per Gibbs sample, as was seen in previous analysis \citep{bp03}. Second to this are the other components grouped into "other TOD processing", which is a upper limit on all the other steps required to sample the LiteBIRD TOD data model. Scaling linearly from the reduced to the full \LB\ data volume, we estimate that a full Gibbs sample will cost 3000\,CPU-hours, which would put the cost of a 1000-sample production run at about 3M CPU-hours.

%% file: Conlusions.tex
\section{Summary and conclusions}
\label{sec:Conclusions}

The main goal of this paper was to assess the computational feasibility of running simulated \LB\ TOD through \commanderthree\ doing iterative end-to-end Bayesian analysis. By modifying \commanderthree\ to fit the \LB\ specific data characteristics, and run it on a partial simulation of what real \LB\ data would look like, we have demonstrated the conceptual feasibility of such an analysis. Furthermore, estimates of the expected final \LB\ data volume compared with the data volume used in this paper shows that such an end-to-end analysis is also computationally feasible. For an analysis of the same data within a traditional simulation-based framework, we refer the interested reader to ref.~\citep{bortolami:2025}.

The analysis for 88 detectors and one year of observations required 700\,GB of RAM and 23\,wall hours per full Gibbs sample on a 128\,core computer node with 2\,TB RAM. We have shown that scaling this up to the full mission will require approximately 70\,TB of RAM with the current implemented compression algorithm and 3000\,CPU hours.
Given the exponential development in computational power, it is likely that a final \LB\ analysis in ten years' time will have access to much larger computational resources, which should be able to achieve the rate of tens of samples/day that we are able to reach for our current analyses.

We have showed that using \commanderthree\ as an analysis pipeline for simulated \LB\ data is feasible. The next step is to implement support for all the \LB\ specific features, from half-wave plate non-idealities and gain fluctuations to beam and sidelobe modelling, and then perform a full end-to-end analysis, with final $r$ estimation as the end goal. 

Based on the current analysis, however, we can now conclude that, with a cost of around 3000\,CPU-hours per Gibbs sample, a Bayesian Cosmoglobe-style analysis of \LB\ is within the capabilities of currently available high-performance computing centers, although with a wall-time that is counted in years rather than days or weeks, as has been the case with previous analyses. A final LiteBIRD analysis will likely have to incorporate additional parameters and algorithms to account for the increased complexity of the final data model. This could increase the final CPU cost by a factor of between 1 and 10, which will become clearer as more development work occurs. Accounting for 10 more years of technological and algorithmic development still to be had, exploiting ever more computing cores and RAM, a full end-to-end Bayesian data analysis of the full \LB\ data set will be feasible. The analysis can be done on a stand-alone data set and it can also include other data sets to take advantage of the Cosmoglobe approach demonstrated in ref.~\citep{watts2023cosmoglobe}. 

In anticipation of the demands of future datasets like \LB, we are currently in the process of developing \texttt{Commander4}, a massively parallelised version of the core Commander algorithms that can be run across a huge number of nodes on a supercomputing cluster. By optimising the distribution of data and computation across many cores, we can greatly improve our abilities to store large datasets in memory as well as process time ordered data in a distributed way. \texttt{Commander4} will relax the memory and computation time constraints present in the \commanderthree\ code, allowing this approach to be distributed over a large number of smaller nodes, instead of the large memory nodes that it is currently optimised for. A main goal of this effort is to perform a full Bayesian end-to-end analysis of \LB\ in the course of a few weeks or months.

%% file: common/Acknowledgments.tex
\section*{Acknowledgments}
%
  The current work has received funding from the
  European Union’s Horizon 2020 research and innovation programme
  under grant agreement numbers 819478 (ERC; Cosmoglobe),
  772253 (ERC; Bits2Cosmology), and 101007633 (MSCA;
  CMBInflate).  Some of the results in this paper have been
  derived using healpy \citep{Zonca_2021} and the HEALPix
  \citep{gorski2005} package. 
This work is supported in Japan by ISAS/JAXA for Pre-Phase A2 studies, by the acceleration program of JAXA research and development directorate, by the World Premier International Research Center Initiative (WPI) of MEXT, by the JSPS Core-to-Core Program of A. Advanced Research Networks, and by JSPS KAKENHI Grant Numbers JP15H05891, JP17H01115, and JP17H01125.
The Canadian contribution is supported by the Canadian Space Agency.
The French \textit{LiteBIRD} phase A contribution is supported by the Centre National d’Etudes Spatiale (CNES), by the Centre National de la Recherche Scientifique (CNRS), and by the Commissariat à l’Energie Atomique (CEA).
The German participation in \textit{LiteBIRD} is supported in part by the Excellence Cluster ORIGINS, which is funded by the Deutsche Forschungsgemeinschaft (DFG, German Research Foundation) under Germany’s Excellence Strategy (Grant No. EXC-2094 - 390783311).
The Italian \textit{LiteBIRD} phase A contribution is supported by the Italian Space Agency (ASI Grants No. 2020-9-HH.0 and 2016-24-H.1-2018), the National Institute for Nuclear Physics (INFN) and the National Institute for Astrophysics (INAF).
Norwegian participation in \textit{LiteBIRD} is supported by the Research Council of Norway (Grant No. 263011 and 351037) and has received funding from the European Research Council (ERC) under the Horizon 2020 Research and Innovation Programme (Grant agreement No. 772253, 819478, and 101141621).
The Spanish \textit{LiteBIRD} phase A contribution is supported by MCIN/AEI/10.13039/501100011033, project refs. PID2019-110610RB-C21, PID2020-120514GB-I00, PID2022-139223OB-C21, PID2023-150398NB-I00 (funded also by European Union NextGenerationEU/PRTR), and by MCIN/CDTI ICTP20210008 (funded also by EU FEDER funds).
Funds that support contributions from Sweden come from the Swedish National Space Agency (SNSA/Rymdstyrelsen) and the Swedish Research Council (Reg. no. 2019-03959).
The UK  \textit{LiteBIRD} contribution is supported by the UK Space Agency under grant reference ST/Y006003/1 - "LiteBIRD UK: A major UK contribution to the LiteBIRD mission - Phase1 (March 25)."
The US contribution is supported by NASA grant no. 80NSSC18K0132.
%
%
%
%